# Comments on Possible Variation of the Universal Constants

A. D. Law and J. Dunning-Davies,
Department of Physics,
University of Hull,
Hull HU6 7RX,
England.

j.dunning-davies@hull.ac.uk

## Abstract.

Discussion of the constancy, or otherwise, of the various so-called 'universal constants' which abound in physics has continued for many years. However, relatively recent observations, which appear to indicate a variation in the value of the fine structure constant, have reignited the subject. These observations relate to quasars and that in itself raises questions. Also, since many of the arguments utilise the Bekenstein-Hawking expression for the entropy of a black hole, further controversy is introduced into the discussion immediately. In what follows these two points will be examined and some less well-known theoretical considerations introduced which, hopefully, will instigate wider examination of this topic.

# Introduction.

The recent revelations by Webb [1] concerning the possible variation of the fine structure constant have led, predictably, to an upsurge of interest in the overall question of the constancy, or otherwise, of the universal constants of physics. It needs to be noted from the beginning that these so-called constants fall neatly into two categories: those which carry dimensions and those which are dimensionless. This is not a pointless differentiation since all systems of units are manmade and so, if a constant has dimensions, it is seemingly linked to something which is manmade and, in that sense, not truly fundamental. However, with the current interest being provoked by possible variation in the value of the fine structure constant, that drawback is eliminated since the fine structure constant is a dimensionless quantity. However, since it is given by the expression

$$\alpha = \frac{e^2}{\hbar c},$$

it is seen to depend on three other quantities, normally regarded as physical constants – the electronic charge $e$, the speed of light $c$, and Planck's constant $\hbar$ – all of which have units associated with them. Hence, if $\alpha$ does vary, any variation will be due to separate variations in one or more of these three quantities. For many years there has been serious speculation that the speed of light is not a constant but, more recently, the possibility of the value of the electronic charge altering has come in for serious consideration. The biggest problem with this latter investigation is that the investigations seem to take the Bekenstein-Hawking expression for the entropy of a black hole as a starting point or, if not as the actual starting point, certainly as a crucial component in the argument. As has been pointed out on numerous occasions, this poses considerable problems when the effect on thermodynamics is examined closely and, crucially, with an open mind devoid of preconceived notions.

Spectroscopic observations of distant gas clouds in conjunction with passing quasars may be used to measure possible variation in the fine structure constant. Variations in $\alpha$ would cause detectable shifts in the rest wavelength of red shifted UV resonance transitions seen in quasar absorption patterns [1]. The initial method conducted by Webb *et al* involved the relativistic fine-structure splitting of alkali doublets (AD). This is useful since the separation between the lines is proportional to $\alpha^2$ and it can be approximated to a good accuracy that any small variations in the separation are directly proportional to $\alpha$ [2]. The AD method offers the advantage of being simple but it fails to capitalise on the data at hand since it compares transitions with reference to one ground state, and only one single doublet is available for analysis each time it is conducted. Another restriction on the potential accuracy is the relativistic effect causing the splitting; this is rather small and hard to detect.

Webb *et al* introduced a new method to combat the problems outlined with the AD method. The new technique introduced what was called the "many-multiplet" (MM) method which was far more sensitive than the previous technique. One crucial aspect of the MM arrangement is that this method considers atoms or ions in a number of states. This is beneficial because the atom will spend less time near the nucleus in an excited state compared to when it features in the ground state. This occurs since $α$ is essentially a measure of the strength between the electron and the nucleus and the comparison between a number of states is vital to give a value regarding a change in $α$ relative to the number of states. This allows more choice instead of concentrating solely on the one ground state.

Another advantage of this method is that lighter elements like magnesium do not react strongly to changes in $α$, but heavier elements such as iron do. Therefore, the lighter elements can be used as "anchors" [1] so that changes in the heavier elements can be more accurately defined.

With this readily at hand, the first proposal was to reanalyse the initial sample in conjunction with new data using the *Keck I telescope* in Hawaii [1], which features an astounding spectral resolution of approximately 7km/s for the entire data set. The old results, now with new definitions of spectral ranges, are composed of 28 Mg/Fe systems over a redshift range of $0.5 < z < 1.8$. The new results feature 13 quasars over a substantially greater redshift range of $1.8 < z < 3.5$ as well as two further absorption systems where the average red shift range is higher than that of the previous methodology. Therefore, Webb *et al* had to incorporate new multiplets such as NiII and CrII which suit the higher ranges.

The question of whether the results obtained may be accounted for by relative systematic errors that are likely to affect the system is worthy of discussion. A wide range of possible sources of systematic effects was considered ranging from kinematic effects to line bleeding. However, Webb *et al* provide two possible experimental effects that could result in the original measurements being undermined; atmospheric dispersion and isotopic abundance evolution [1]. The first concerns light dispersion from quasars, since light passing through the earth's atmosphere is subject to dispersion depending from which part of the frequency spectrum it originates. If the direction of the differential dispersion and spectrograph are not aligned, a change would be incurred in the ratios, which incidentally would enhance them by pushing the ratios closer to the positive end of the spectrum.

Similarly, the same occurs when the error regarding isotopic abundance is examined. This effect concerns the relative amounts of different isotopes in the same elements in the quasar spectra and laboratory spectra [1]. It is likely that solar abundances are different from terrestrial values, so a method of removing weaker isotopes in all the relevant samples and recalculating the alpha ratios can account for the effect of again pushing the alpha ratio to more positive values. Thus, Webb concludes stating that *"applying either of the two significant corrections would enhance the significance of our results"* [1]. This seems a very good step regarding the validity of the data offered. However it does appear to be necessary for a full independent survey of these corrections to be conducted.

Webb's results refer to situations concerning quasars with certain red shift values. However, the first worry in any such discussion surrounds the true meaning of any obtained red shift value. The true meaning of any red shift value is still a matter under discussion in some quarters and is not something which can be glossed over in any serious scientific examination. The work of Halton Arp [3] on the red shift of quasars cannot sensibly be dismissed and it might be noted usefully at this point that some recent thoughts on a varying speed of light lend credence to Arp's theories [4,5].

**Varying speed of light theories.**

The idea that the speed of light is a constant seems to have been regarded as an almost self-evident fact since Einstein's Special Theory of Relativity became a widely accepted scientific truth. However, Einstein himself assumed the speed of light in a vacuum constant; he did not assume the speed of light itself constant. This is a much more acceptable assumption since it is known from experiment that the speed of light is definitely not a constant; its value depends on the medium through which it is passing. This has led Santilli [4] to speculate that the speed of light depends on the refractive index of the material through which it is passing and it is by using Santilli's basic theory that Mignani [5] has been able to explain Arp's interesting observations of quasars – observations that many attempt to dismiss since they bring into question the accepted explanation of the red shift. By assuming the speed of light depends on the refractive index of the medium through the light is passing, a more acceptable interpretation of red shift is achieved and it is one which allows for Arp's observations.

However, in 1985, Thornhill [6] showed that the speed of light should vary with the square root of the background temperature. This result means that the speed of light will vary with the passing of time and, incidentally, does away with a need for a period of inflation in the early universe if a big bang scenario is accepted. This latter point re-emerged later in the work of Moffat [7] and Albrecht and Magueijo [8]. Hence, once the notion of a speed of light varying with the passage of time is accepted, the possibility of the fine structure constant varying also through its dependence on the speed of light arises. Obviously, however, since the fine structure constant depends on other constants also, if they vary also, the various variations could cancel one another or could combine to enhance each others contribution.

**Varying electronic charge theories.**

The fundamental notion of a varying value for the electronic charge generally stems from the accepted ideas associated with black holes and black hole thermodynamics. Hence, the notion immediately runs into problems. Firstly, the modern idea of a black hole is not associated with Michell's idea stemming from 1784 and based purely on Newtonian mechanics [9] of a body which possesses an escape speed greater than that of light. The modern idea is related to a singularity found in the so-called Schwartzschild solution of Einstein's field equations of General Relativity. The first major problem encountered here is that the popular expression for this so-called Schwartzschild solution is not the expression appearing in Schwartzschild's original paper [10]. In Schwartzschild's original paper, no such singularity appears. This immediately raises grave doubts about the modern idea of a black hole.

Further doubts arise when the currently accepted version of so-called black hole thermodynamics is considered. Given the well-known thermodynamic result that the entropy of a system never decreases, it was almost inevitable that the result of Hawking's area theorem for a black hole would be linked with the notion of entropy of a black hole since he showed that the area of a black hole would not decrease. Bekenstein duly made the connection and it is now accepted by many that the entropy of a black hole is given by the Bekenstein-Hawking expression

$$S_{bh} = \frac{kM^2}{\sigma_m^2},$$

where $M$ is the 'irreducible' mass of the black hole and $\sigma_m = (ch/2\pi G)^{1/2} = 2 \times 10^5$ gm is the Planck mass. This entropy expression refers, of course, to an uncharged, non-rotating black hole. For a charged black hole, the $M^2$ is replaced by $\left(M + \sqrt{\left(M^2 - Q^2/G\right)}\right)^2$. However, whichever expression is used, the same problem occurs. The accepted black hole entropy expression is neither extensive nor concave. Hence, because of its first deficiency, many commonly used thermodynamic expressions, such as the Euler relation and the Gibbs-Duhem equation are no longer available for use. The second deficiency has been shown to lead to the possibility of violations of the Second Law of Thermodynamics. It would be perfectly reasonable to discuss the possibility of the Second Law being violated; it is, after all, merely a fact of experience. However, it is a fact of experience which has stood the test of time throughout a wide range of examples and its possible violation should be viewed openly and with great caution, - particularly in this case where the entire modern notion of a black hole is open to question. It is not, after all, normally regarded as acceptable to impose a physical interpretation on a mathematical singularity but, in essence, that is precisely what is done here.

The end result of this has to be to raise grave doubts about any arguments based on the modern theory of black holes and particularly on the thermodynamics of those bodies. It might be noted in passing that it is not absolutely necessary for a star too massive to form a neutron star as its end-point to result in collapse to a black hole; the possibility of quark stars and even sub-quark stars could obviate this [11].

**Conclusions.**

Whether or not the values of the various constants of physics do, in fact, change over time is still very much an open question. As far as the fine structure constant is concerned, there appears to be some evidence suggesting that it is not actually a constant. However, it is a combination of other quantities normally regarded as constant. It is undoubtedly true that one of these, the speed of light, is not a true constant and, as has been mentioned, quite a lot of work has been done examining this and considering the consequences that follow. Another, the value of the charge on an electron, has been examined but since the investigation has been so dependent on the presently accepted expression for the entropy of a black hole, that work must remain open to question. As has been pointed out previously [12], there is serious doubt concerning the validity of those investigations. The status of Planck's constant in all of this remains unaffected so far. However, that constant, together with all the others such as Boltzmann's constant and the universal constant of gravitation, must come under scrutiny if this whole question of the constancy, or otherwise, of the so-called universal constants of physics is to be clarified satisfactorily.

**References.**

[1]  J. K. Webb et al; 2001, Phys. Rev. Lett. **87**, 091301

[2]  J. K. Webb et al; 1998, Phys. Rev. Lett. **82**, 884

[3]  H. Arp; 1998, *Seeing Red*,
     (Apeiron, Montreal)

[4]  R. M. Santilli; 2007, reprint Institute for Basic Research IBR-TPH-03-07

[5]  R. Mignani; 1992, Physics Essays **5**, 531

[6]  C. K. Thornhill; 1985, Speculations Sci. Tech. **8**, 263

[7]  J. W. Moffat; 1993, Int. J. Mod. Phys. D **2**, 351

[8]  A. Albrecht and J. Magueijo; 1999, Phys. Rev. D **59**, 043516

[9]  J. Michell; 1784, Phil. Trans. R. Soc., **74**, 35

[10] K. Schwartzschild; 1916, Sitzungsberichte der Königlich Preussischen
      Akademie der Wissenschaften zu Berlin, Phys-Math. Klasse, 189

[11] G.H.A.Cole & J.Dunning-Davies; 1999, Gravitation, **4**, 79

[12] J. Dunning-Davies: 2003, in *Focus on Astrophysics Research*,
      ed. Louis V. Ross, (Nova, New York)